\documentclass[prl,twocolumn,showpacs,superscriptaddress]{revtex4}
\usepackage{amsmath,amssymb}
\usepackage{graphicx,color}
\usepackage{times}

\begin{document}

\newcommand \be  {\begin{equation}}
\newcommand \bea {\begin{eqnarray} \nonumber }
\newcommand \ee  {\end{equation}}
\newcommand \eea {\end{eqnarray}}

\title{Freezing Transition, Characteristic Polynomials of Random Matrices, $ \hspace{5cm} \mbox{and}$ the Riemann Zeta-Function}
\author{Yan V. Fyodorov}
 \affiliation{Queen Mary University of London, School of Mathematical Sciences, London E1 4NS, United Kingdom}
 \author{Ghaith A. Hiary} \author{Jonathan P. Keating}
 \affiliation{School of Mathematics,
University of Bristol, Bristol BS8 1TW, UK}
%\received{}
\published{{\textbf{Phys. Rev. Lett. 108 , 170601 (2012)}} }

\begin{abstract}
We argue that the {\em freezing transition scenario}, previously
explored in the statistical mechanics of $1/f-$noise random energy models,
also determines the value distribution of the maximum of the modulus of the characteristic polynomials of large
$N\times N$ random unitary (CUE) matrices.  We postulate that our results extend to the extreme values taken by the Riemann zeta-function $\zeta(s)$ over sections of the critical line $s=1/2+it$ of constant length and present the results of numerical computations in support.  Our main purpose is to draw attention to possible connections between the statistical mechanics of random energy landscapes, random matrix theory, and the theory of the Riemann zeta function.
\end{abstract}
\maketitle
The Riemann zeta-function
\be\label{zetadef}
\zeta(s)=\sum_{n=1}^{\infty}\frac{1}{n^s}=\prod_{p}\left(1-\frac{1}{p^s}\right)^{-1}
\ee
encodes the distribution of the primes $p$ in the positions of its {\em non-trivial zeros} \cite{Titchmarsh}.  The Riemann Hypothesis, which is of central importance in mathematics, places these zeros on the critical line $s=1/2+it, \, t\in \mathbb{R}$.  There has in recent years been considerable interest, stemming from speculations about a spectral interpretation of the Riemann zeros,  in possible connections between the theory of the zeta-function and quantum mechanics (see, e.g.~\cite{qm}).  Our focus here is on suggesting a new link with physics via the statistical mechanics of disordered landscapes.

Some of the most significant questions in the theory of the zeta function concern the distribution of values it takes on the critical line, where it behaves like a quasi-random function of $t$
(essentially because the phases of the terms in the product in (\ref{zetadef}) contribute as if they were random)
.  It was proved by Selberg, for example, that, as $t\rightarrow\infty$, $\log |\zeta(1/2+it)|$
behaves like a Gaussian random variable with mean zero and variance $\frac{1}{2}\log\log\frac{t}{2\pi}$, see e.g.~\cite{Lau}.
In other words, the typical size of $\log |\zeta(1/2+it)|$ is of the order of $\sqrt{\log\log t}$.  As regards the extreme values taken by the zeta function over long ranges, the Lindel\"of Hypothesis asserts that $|\zeta(1/2+it)|$ grows more slowly than any power of $t$ as $t \rightarrow\infty$, the Riemann Hypothesis implies that it is $O(\exp ({\rm const.} f(t)))$, where $f(t)=\log t/\log\log t$, while it is known that $|\zeta(1/2+it)|$ takes values at least as large as $\exp (\sqrt{f(t)})$ infinitely often \cite[pp.~354,~209]{Titchmarsh}.  The extreme values thus lie in the range between these upper and lower bounds (and are significantly larger than the typical values).  The problem of determining where precisely within this range they lie has attracted considerable attention, but remains unresolved.  The extreme values in question are so rare that numerical computations have failed to settle the matter so far.

One model, proposed by Montgomery, that leads to predictions for the extreme values is based on the assumption that the local maxima of $\log|\zeta(1/2+it)|$ are statistically independent and that they obey the normal distribution proved for general values. Under such assumptions one finds that the typical size of the maximum value of $|\zeta(1/2+it)|$ should be of the order of $\exp\left(c_M \sqrt{\log (t) \log\log (t)}\right)$ \cite{FGH}, where $c_M$ is a constant.  This is closer to the lower bound than the upper bound, implying that the extreme values are not much larger than the largest value known to be reached infinitely often.

However, the values of $\log|\zeta(1/2+it)|$ are, in fact, strongly correlated in a way that is significant from the point of view we shall  explore here. Specifically, define, for a fixed $t\in \mathbb{R}$,
\be\label{logzetaint}
 V_t^{(\zeta)}(x)=-2\log{|\zeta\left(\frac{1}{2}+i(t+x)\right)|}
\ee
Selberg's theorem  implies that $V_t^{(\zeta)}(x)$
behaves like a Gaussian random function of  $x$ when $t\to\infty$. Such a random process is characterized by the two-point correlation function $C(x_1,x_2)=\overline{V^{(\zeta)}_t(x_1)V^{(\zeta)}_t(x_2)}$, with the bar denoting the average over intervals $[t-h/2,t+h/2]$, with $h\ll t$ chosen so that the intervals contain, asymptotically (as $t\to\infty$), an increasing number of zeros.
A straightforward calculation, sketched in \cite{FyoKeat}, see also \cite{Bourg}, leads to
\be\label{corzetadiag3int}
 C(x_1,x_2)=\left\{\begin{array}{c} -2\log{|x_1-x_2|}, \,\, \mbox{for} \quad  \frac{1}{\log{t}}\ll |x_1-x_2|\ll 1\\
2\log{\log{t}}, \,\, \mbox{for} \quad  |x_1-x_2| \ll \frac{1}{\log{t}} \end{array}\right.
\ee
The significance of the {\it logarithmic} form of the correlations will become apparent when we come to make comparisons with corresponding problems in random matrix theory and statistical mechanics.

Over the past 40 years, inspired by the pioneering work of Montgomery \cite{Montpair} and Odlyzko \cite{Odl}, considerable evidence has accumulated for connections between the statistical properties of the Riemann zeta-function and those of large random matrices.   For example, correlations between the nontrivial zeros of the zeta function are believed to coincide asymptotically with those between the eigenvalues of large random unitary or hermitian matrices \cite{Montpair,Odl}, and the value distribution of $\zeta(1/2+it)$ is believed to be related to that of the characteristic polynomials of these matrices  \cite{KeatSna00,HKO,CFKMS,GHK}.  Our purpose here is to connect these two areas of research to a third, the statistical mechanics of disordered logarithmically-correlated energy landscapes, see \cite{CLD,FB,FLDR} and references therein.  The analogy we develop suggests that the spin-glass-like freezing transition that dominates the low-temperature behaviour in the statistical mechanical problems also governs the extreme values taken by the characteristic polynomials of random matrices and the zeta function.  This sheds new light on the longstanding problem of determining the maximum size of
%the zeta function.
$\zeta(1/2+it)$.

To establish such a connection we consider the ensemble of $N\times N$ unitary matrices, chosen uniformly at random from the unitary group ${\cal U}(N)$ (i.e.~from the Circular Unitary Ensemble, or CUE).  We denote the eigenvalues of a given matrix $U_N$ by $\exp{(i\phi_1)},\ldots, \exp{(i\phi_N)}$ and the corresponding characteristic polynomial by
\be\label{pdef}
p_N(\theta)=\det{\left(1-U_N\,e^{-i\theta}\right)}=\prod_{n=1}^{N}\left(1-e^{-i(\phi_n-\theta)}\right).
\ee
It is instructive to compare $V_t^{(\zeta)}(x)$ from (\ref{logzetaint}) with  $V^{(U)}_N(\theta)=-2\log{|p_N(\theta)|}$. It was proved in \cite{KeatSna00} that $V^{(U)}_N(\theta)$ satisfies a central limit theorem: the distribution of the values of $\log|p_N(\theta)|$ tends to a normal with mean zero and variance $2\log{N}$ as $N\rightarrow\infty$.  Identifying the mean density of the eigenvalues, $N/2\pi$, with the mean density of the Riemann zeros near to height $t$,
$\frac{1}{2\pi}\log\frac{t}{2\pi}$, renders the agreement with Selberg's theorem for $\log |\zeta(1/2+it)|$ complete.  Importantly for us here, $V^{(U)}_N$ has the following representation \cite{HKO}:
 \be\label{8int}
V^{(U)}_N(\theta)=\sum_{n=1}^{\infty}\frac{1}{\sqrt{n}}\left[e^{-i n\theta}v^{(N)}_n+\mbox{comp. conj.}\right]
\ee
with $\sqrt{n}v^{(N)}_n=\mbox{Tr}\left(U_N^{-n}\right))$.
According to \cite{DiaSha}, as $U_N$ varies in the CUE the coefficients $v^{(N)}_n$ for any fixed $n$ tend in the limit $N\to \infty$ to
 independent, identically distributed complex gaussian variables with zero mean and unit variance. Denoting averages over the unitary group ${\cal U}(N)$ with the angular brackets, a simple calculation  shows that
$\left\langle V^{(U)}_N(\theta_1)V^{(U)}_N(\theta_2)\right\rangle$ tends in the limit $N\to \infty$ to
 $-2\log\left(2|\sin\frac{1}{2}(\theta_1-\theta_2)|\right)$, and so exhibits precisely the same logarithmic behaviour at small distances as in (\ref{corzetadiag3int}).  For large but finite $N$, the logarithmic divergence can be shown to saturate at $|\theta_1-\theta_2|\sim N^{-1}$, so after associating $N\sim \log \frac{t}{2\pi}$ the correspondence between the small-scale behaviour of $V^{(U)}_N(\theta)$ and $V_t^{(\zeta)}(x)$ becomes complete.  This is significant from the point of view we seek to develop.

Our primary goal is to determine the distribution of the maximum value of $|p_N(\theta)|$ over  $0\le\theta\le 2\pi$ when the matrix $U_N $ ranges over ${\cal U}(N)$.  Our second goal is then to use the random-matrix results to motivate predictions for the extreme values of the Riemann zeta function.   The first steps in this direction were taken in \cite{FGH}, where the tail of the distribution of the maximum values of $|p_N(\theta)|$ in $0\le\theta\le 2\pi$ was found.  This tail determines the typical size of the maximum values expected when  $U_N $ is sampled independently a large (exponentially in $N$) number of times from within ${\cal U}(N)$.  The results are consistent with the Montgomery model and lead to a prediction for the value of the constant $c_M$ there.

The focus here will differ from that of \cite{FGH} in the following ways.  We shall be concerned with the distribution of maximum values of the characteristic polynomials of single matrices, rather than with large numbers of matrices, and will obtain the full value distribution of the maxima in the limit as $N\rightarrow \infty$, rather than concentrating on the tail that is relevant when maximizing over many matrices.  This leads to a model for the distribution of maximum values of $|\zeta(1/2+it)|$ over the intervals $T \le t \le T+2\pi$,  rather than $0 \le t \le T$ as $T\rightarrow\infty$.  Furthermore, it makes the problem of numerical computation of the distribution in question significantly easier, because one is finding the maximum of $\sim \log T$ rather than $\sim T \log T$ numbers (the values of the local maxima).

We start by noting that the maximum value of  $p_N(\theta)$ can be characterized in terms of
\be\label{1}
{\cal Z}_{N}(\beta)=\frac{N}{2\pi}\int_0^{2\pi}|p_N(\theta)|^{2\beta}d\theta\equiv \frac{N}{2\pi}\int_0^{2\pi}e^{-\beta V_N(\theta)}\,d\theta,
\ee
where $V_N(\theta)=-2\log{|p_N(\theta)|}$ and $\beta>0$.  Specifically, if ${\cal F}(\beta)=-\beta^{-1}\log{{\cal Z}_N(\beta)}$, then
\be\label{2}
\lim_{\beta\to \infty}{\cal  F}(\beta)=\min_{\theta\in(0,2\pi)}V_N(\theta)=-2\max_{\theta\in(0,2\pi)}
 \log{|p_N(\theta)|}.
\ee
The key observation is that (\ref{1}) takes the form of the partition function for a system with energy
$V_N(\theta)$ and inverse temperature $\beta$.  ${\cal F}(\beta)$ may then be associated with the corresponding free energy.  Recalling that the values of $V_N(\theta)$ are asymptotically gaussian distributed and logarithmically correlated, it is natural to draw comparisons with a class of problems that has attracted a good deal of attention recently in the area of disordered systems, namely the statistical mechanics of a particle equilibrated in a random landscape with logarithmic correlations. Two-dimensional systems of that sort appear in many contexts, for example
in the problem of Dirac fermions in a random magnetic field \cite{CMW}, and one-dimensional analogues were considered recently in  \cite{FB,FLDR}.  Importantly for us, in the one-dimensional case the energy landscape is given by a regularization of the random Fourier series featuring in (\ref{8int}).  In the statistical mechanical problem there has been a particular focus on the so-called freezing transition \cite{CMW,CLD} which dominates the low temperature limit and determines the extreme value statistics. The latter appears to be manifestly different from the case of uncorrelated variables.  We shall argue that a similar freezing transition determines the extreme value statistics of the characteristic polynomials and hence, conjecturally, of $|\zeta(1/2+it)|$, and that therefore the logarithmic correlations exhibited by $V^{(U)}_N(\theta)$ and $V_t^{(\zeta)}(x)$ play an important role.

The first step is to consider the positive integer moments $\left\langle{\cal Z}_N^k(\beta)\right\rangle, \quad k=1,2,\ldots$.
Using standard methods of random matrix theory (RMT) it is easy to show that
\begin{eqnarray}\label{3}
\left\langle{\cal Z}_N^k(\beta)\right\rangle = N^k\int_0^{2\pi}\ldots \int_0^{2\pi} && \frac{D_N^{(k)}(\beta)}{D_N^{(k)}(0)}\prod_{j=1}^k\frac{d\theta_j}{2\pi},\,
\end{eqnarray}
  where $D_N^{(k)}(\beta)=\det{\left(M_{i-j}^{(\beta)}\right)_{i,j=0}^{N-1}}$ is the determinant of a Toeplitz matrix
\be\label{5}
M_{i-j}^{(\beta)}=\int_0^{2\pi} e^{i\phi (i-j)}\prod_{p=1}^{k}\left[2-2\cos{(\phi-\theta_p)}\right]^{\beta}\frac{d\phi}{2\pi}\,.
\ee
When $N\to \infty$ the asymptotics of such Toeplitz determinants is well known \cite{Wid} to be given by
 \be\label{6}
D_N^{(k)}\approx \left[N^{\beta^2}\frac{G^2(1+\beta)}{G(1+2\beta)}\right]^k \prod_{r<s}^k|e^{i\theta_r}-e^{i\theta_s}|^{-2\beta^2}\,,
\ee
where $G(x)$ is the so-called Barnes function. Substituting (\ref{6}) back to  (\ref{3}) we see that the resulting expression is the standard Dyson-Morris
version of the Selberg integral \cite{FWSel},
 convergent for $k<\frac{1}{\beta^2}$, and divergent for larger $k$. As we have $k\ge 1$ the procedure makes sense only for $\beta^2~<~1$.
 Introducing ${\cal Z}_e=N^{1+\beta^2}\frac{G^2(1+\beta)}{G(1+2\beta)\Gamma(1-\beta^2)}$, we find
 \be\label{7}
\left\langle{\cal Z}_N^k(\beta)\right\rangle= {\cal Z}_e^k \Gamma(1-k\beta^2), \quad k<\beta^{-2}\,.
\ee

The expression (\ref{7}) for the moments has exactly the same form as that for the partition function of the landscape of
 the circular-logarithmic model (a periodic version of $1/f$ noise) \cite{FB}, but with a different value for
 the characteristic scale ${\cal Z}_e$.  This means we can simply translate, {\it mutatis mutandis}, the results of \cite{FB}
 to the values of characteristic polynomial sampled along the full circle $\theta\in(0,2\pi]$. In particular, we conclude that
  the maximum value of the modulus of a CUE characteristic polynomial $p_N(\theta)$ in an interval $\theta\in[0, 2\pi]$ can be written in the limit $N\rightarrow \infty$ as
\be\label{12sum}
-2\max_{\theta\in[0,2\pi)}{\log{|p_N(\theta)|}}\sim a_{N}+b_{N}\,x,
\ee
where $a_{N}=-2\log{N}+c\log{\log{N}}+o(1)$, with, conjecturally, $c=\frac{3}{2}$, $b_{N}=1+O(1/\log{N})$, and $x$  is a random variable taking values distributed with probability density
\begin{equation}\label{prediction1}
p(x)=
%-g_{\beta_c}'(x)=-\frac{d}{dx}\left[2 e^{x/2} K_1(2 e^{x/2})\right]=
2e^{x}K_0(2 e^{x/2}),
\end{equation}
where $K_{\nu}(z)$ denotes the modified Bessel function. Two consequences are particularly noteworthy: (i) the tail of the probability density of the random variable $x$ is given asymptotically when $x\to -\infty$ by $p(x)\sim |x|e^{x}$, (ii) the value  of the constant $c=3/2$. These two features are believed to be {\it universal characteristics} of the extreme value statistics of the class of logarithmically correlated random variables \cite{CLD}, and distinguish those from short-range correlated random variables, for which $c=1/2$ and $p(x)\sim e^{x}$ (Gumbel distribution) \cite{LLR}. This new class is believed to include, in particular, the $2D$ Gaussian free field \cite{BZ}, branching random walks \cite{ABK}, polymers on disordered trees \cite{DS}, $1/f$ noise \cite{FB,FLDR} and models appearing in turbulence and financial mathematics \cite{turbfin}.

In order to illustrate the extreme value predictions in the context of the Riemann
zeta function we now summarize the results of preliminary numerical computations.  These involved numerically evaluating $\zeta(1/2+it)$ over ranges of length $2\pi$, at various heights $T$, and finding the maximum value $\zeta_{\rm max}(2\pi; T)$ in each range.
%Values of $\zeta(1/2+it)$ were computed
We used the amortized-complexity algorithm \cite{Hia}, which is suitable for computing $\zeta(1/2 + it)$ at many points.
%Point-wise values of $\zeta(1/2+it)$ that we computed are typically accurate to within $\pm 5\times 10^{-11}$, which is sufficient for the purposes of this experiment.
%The maximum of $|\zeta(1/2 + it)|$ between
% consecutive zeros was computed to within $\pm 10^{-9}$.

The first test concerns the value of the constant $c$ in  (\ref{12sum}).  We expect the logarithmic correlations to lead to $c=\frac{3}{2}$, rather than  $c=\frac{1}{2}$, as would be the case if the zeta correlations were short-range. The mean of $\zeta_{\rm max}(2\pi; T)$ suggested by the model in \eqref{12sum} is $\delta = e^{\gamma} N/(\log N)^{\frac{c}{2}}$, with $c = 1/2 \textrm{ or } 3/2\,,$ and $\gamma = 0.57721\ldots\,,$ where we set $N$ to be the nearest integer to $\log T$.
At each height a sample that spans $\approx 10^7$ zeros is used yielding $\approx 10^7/N$ sample points
(since there are roughly $N$ zeros in each range of length $2\pi$).
%$\approx 10^7/\lceil\log{T}\rceil$ sample points.
%given by $\delta = e^{\gamma} \lceil \log{T}\rceil/(\log{\lceil \log{T}\rceil})^{\frac{c}{2}}$,
%and where $\lceil\log{T}\rceil$ is the nearest integer to $\log T$, and $\gamma = 0.57721\ldots$.
%The data are summarized in the table below.
In view of the table below, one may conclude that $c = 3/2$ fits the data considerably better than $c = 1/2$, thus supporting the
logarithmic correlations model.
\begin{table}[ht]
\footnotesize
\caption{\footnotesize Ratio of data mean $\tilde{\delta}$ to model mean $\delta$ with $c = 3/2$ and $c = 1/2$.}\label{mean ratios}
\renewcommand\arraystretch{1.5}
\begin{tabular}{c|c|c|c}
$T$       & $ \qquad N \qquad $        &   $\left(\tilde{\delta}/ \delta\right)_{c=3/2}$ & $\left(\tilde{\delta}/ \delta\right)_{c=1/2}$\\
\hline
$10^{22}$ & 51  &     1.001343    &   0.504993\\
$10^{19}$ & 44  &    0.992672   & 0.510293    \\
$10^{15}$ & 35  &    0.976830    & 0.518057    \\
$3.6\times 10^{7}$ & 17  &  0.930533 &  0.552856
\end{tabular}
\end{table}

\begin{figure}[t!]
{\small }
\includegraphics[width=0.42\textwidth, height=0.26\textheight]{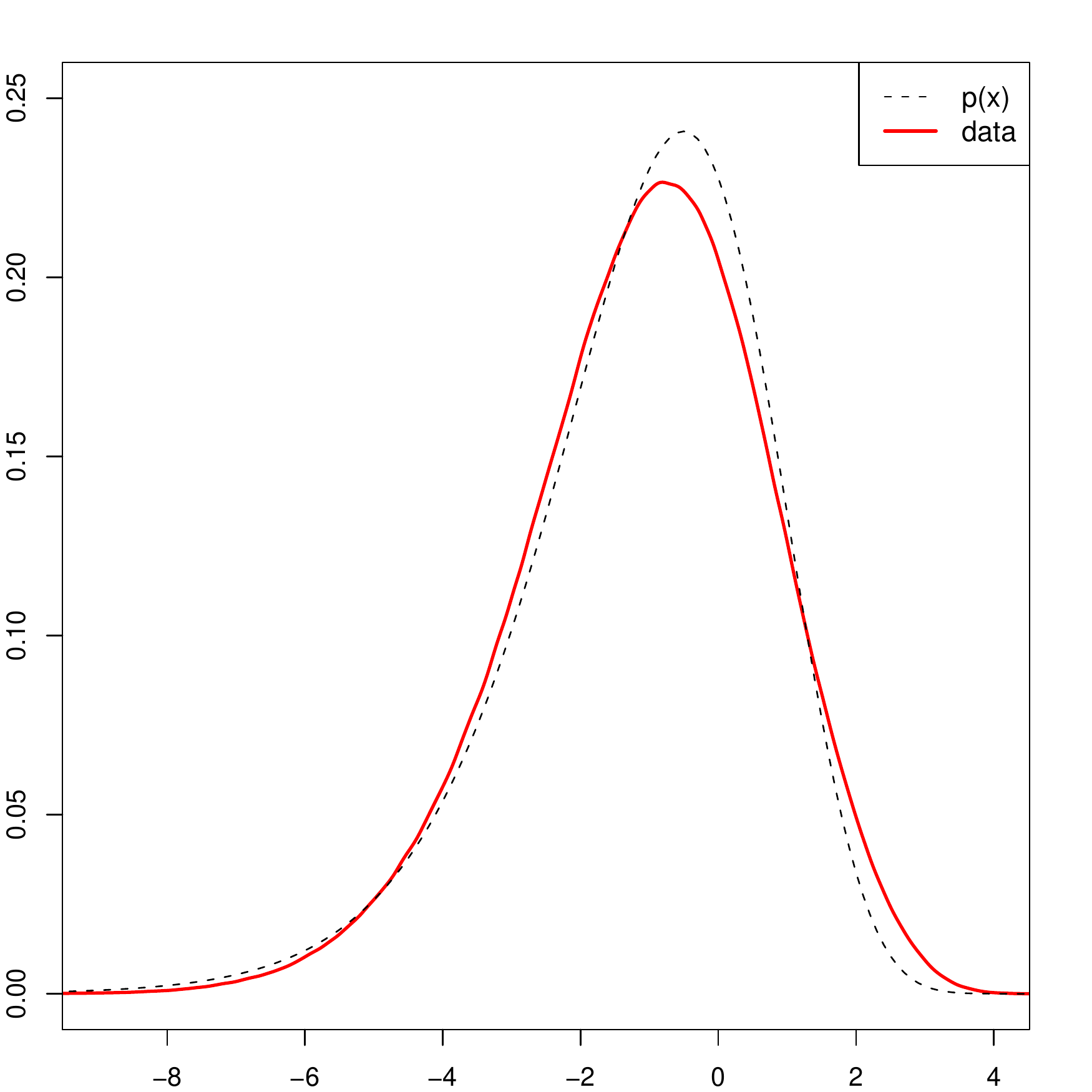}
\caption{Numerical computation (solid red line) compared to theoretical prediction (\ref{prediction1}) (dashed black line) for $p(x)$.}\label{figure 1}
\end{figure}

Testing the distribution $p(x)$ is more difficult, because the data converge extremely slowly at that scale.  The results of our initial experiments are summarized in Figure~\ref{figure 1}.  Specifically, we consider
$ -2 \log |\zeta_{\rm max}(2\pi; T)| + 2\log N - \frac{3}{2} \log\log N$ based on a set of approximately $2.5 \times 10^8$ zeros
near $T = 10^{28}$. The data are normalized so that $ -2 \log|\zeta_{\rm max}(2\pi; T)| + 2\log N - \frac{3}{2} \log\log N$ has empirical variance
$= \int x^2 p(x)\,dx =3.28986813\ldots$.  The overall agreement is supportive of (\ref{prediction1}), especially in the important tail when $x\to -\infty$ and in view of the fact that lower order arithmetical terms \cite{KeatSna00, CFKMS} have not been incorporated, but cannot be said to be a conclusive verification at this stage.  The behaviour in the tail is significant because if it were to persist into the large deviation regime it would suggest that the Montgomery heuristic significantly underestimates the maximum values achieved by $|\zeta(1/2+it)|$, although this seems unlikely.

Finally, we put the specific idea of the freezing to a numerical test by considering the analogue of the partition function ${\cal Z}_{N}(\beta)$ defined by (\ref{1}) which is clearly
\be\label{zeta1}
z_{\beta}(T)=\frac{N_T}{2\pi}\int_T^{T+2\pi}|\zeta(1/2+iy)|^{2\beta}dy,
%\quad N_T=\log\frac{t}{2\pi}
\ee
where $N_T=\log\frac{T}{2\pi}$, and calculating the ensuing free energy $f(\beta)=-(\beta\log{(N_T)})^{-1}\overline{\log{z_{\beta}}}$ by averaging over $10^{6}$  values of $T$ near $T=10^{28}$.
If freezing is operative, -$f(\beta)$  is expected to be equal to $\beta+1/\beta$ for $\beta<\beta_c=1$ and remain frozen to $-f(\beta)=2$
for all $\beta>1$ \cite{CLD,FB,FLDR}. The results shown in Figure 2 (which are renormalized by the subtraction of lower order arithmetic terms \cite{FyoKeat}) again are consistent with this prediction.

\begin{figure}[t!]
{\small }
\includegraphics[width=0.4\textwidth, height=.26\textheight]{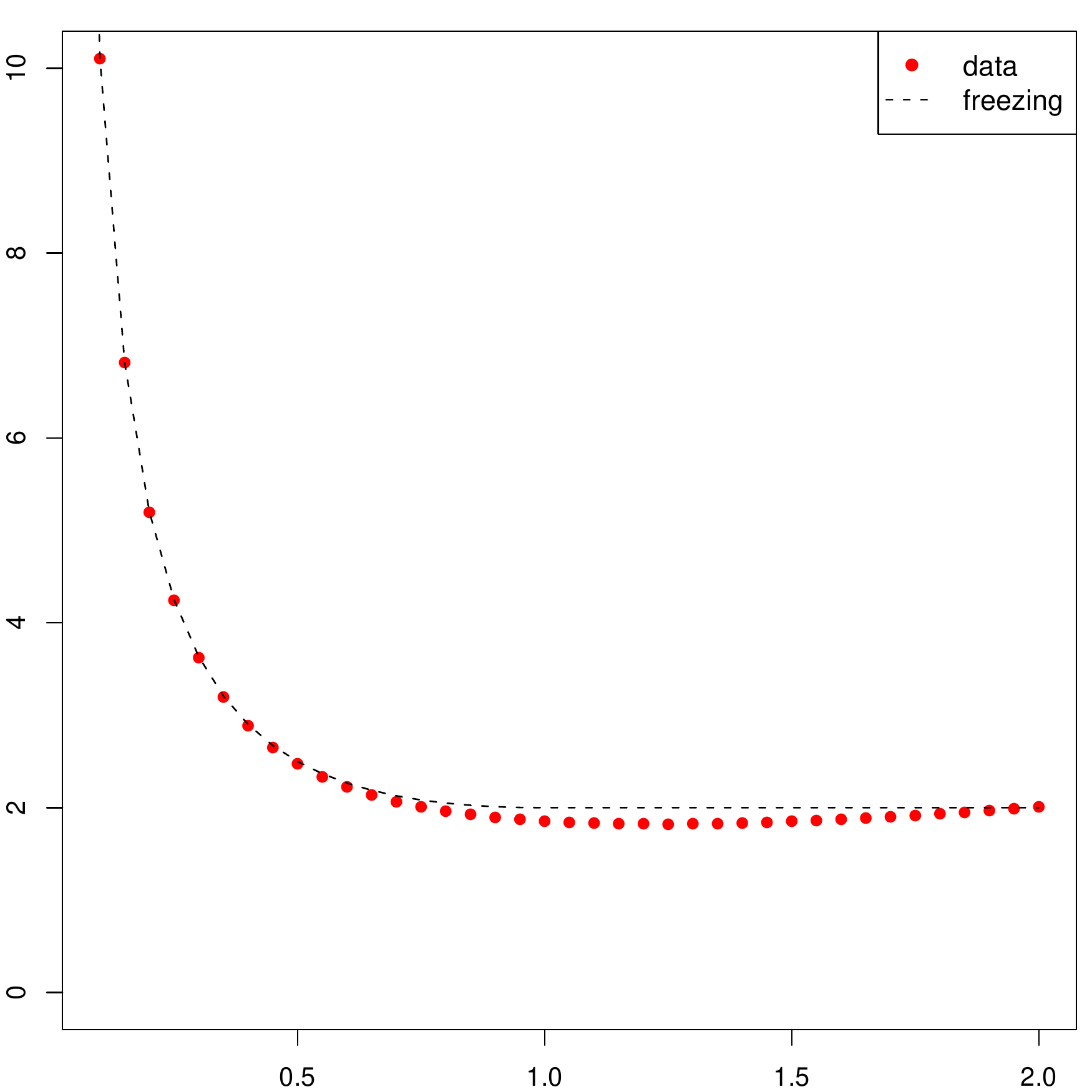}
\caption{Numerical computation (red dots) compared to the theoretical prediction (dashed black line) for  $-f(\beta)$, suggesting freezing beyond $\beta=1$}
\end{figure}

To conclude, we have put forward speculations  on the existence and implications of the freezing phenomena seen in the statistical mechanics of random energy landscape models in the context of the extreme value distribution of the characteristic polynomials of large random matrices  and, further, of the Riemann zeta-function on the critical line.  We believe this sheds interesting new light on these problems.  In a forthcoming paper \cite{FyoKeat} we present the details of our calculations and discuss further the broader picture of freezing phenomena and their manifestation in the present context.

 Support by the EPSRC grant EP/J002763/1 (YVF), the Leverhulme Trust and the AFOSR (GAH and JPK) is gratefully acknowledged.


\begin{thebibliography}{99}%

\bibitem{Titchmarsh} E.C. Titchmarsh {\it The Theory of the Riemann Zeta-function}.  Second Edition, OUP, 1986.

\bibitem{qm} M.V. Berry and J.P. Keating {\it SIAM Review} {\bf 41}, 236 (1999); G. Sierra and P.K. Townsend {\it Phys. Rev. Lett.} {\bf 101}, 110201 (2008); G. Sierra and J. Rodriguez-Laguna {\it Phys. Rev. Lett.} {\bf 106}, 200201 (2011); M. Srednicki {\it Phys. Rev. Lett.} {\bf 107}, 100201 (2011); M.V. Berry and J.P. Keating {\it J. Phys.} A {\bf 44}, 285203 (2011)

\bibitem{Lau} A. Laurincicas {\it Limit Theorems for the Riemann Zeta-Function}. Dodrecht: Kluwer Academic Publishers, 1996

\bibitem{FGH} D.W. Farmer, S.M. Gonek, and C.P. Hughes
{\it J. Reine Angew. Math (Crelle's Journal)}, {\bf 609} 215 (2007)

\bibitem{FyoKeat} Y.V. Fyodorov and J.P. Keating  {\it Freezing Transitions and Extreme Values: Random Matrix Theory, $\zeta(1/2+it)$, and Disordered Landscapes}, in preparation

 \bibitem{Bourg} P. Bourgade, {\it Prob. Theor. Rel. Fields} {\bf 148}, 479 (2010)

\bibitem{Montpair} H. Montgomery in: {\it Proc. Sympos. Pure Math. vol. XXIV, St. Lois. Mo. 1972} Amer. Math. Soc., Providence, R.I. (1973)

\bibitem{Odl} A.M. Odlyzko, The $10^{20th}$ zero of the Riemann zeta function and 70 million of its neighbours, Preprint 1989, unpublished

\bibitem{KeatSna00} J.P Keating and N.C. Snaith {\it Comm. Math. Phys.} {\bf 214} (2000) 57

\bibitem{HKO} C.P. Hughes, J.P. Keating, and  N. O'Connell
 {\it Commun. Math. Phys. }{\bf 220}, 429 (2001)

 \bibitem{CFKMS} J.B. Conrey et al. {\it Commun. Math. Phys.} {\bf 237}, 365 (2003); {\it Proc. London. Math. Soc.} {\bf 91}, 33 (2005);  {\it J. Number Theory} {\bf 128}, 1516 (2008)

 \bibitem{GHK} S.M. Gonek, C.P. Hughes, and J.P. Keating  {\it Duke Math. J.} {\bf 136}, 507 (2007)

\bibitem{CLD} D. Carpentier and P. Le Doussal   {\it Phys. Rev. E} {\bf 63}, 026110 (2001); L.-P. Arguin and O. Zindy, ArXiv: 1203.4216

\bibitem{FB} Y.V. Fyodorov and J.P. Bouchaud  {\it J. Phys. A: Math. Theor.} {\bf
41} 372001 (2008)

\bibitem{FLDR} Y.V. Fyodorov, P. Le Doussal, and A Rosso {\it J. Stat. Mech.},  P10005 (2009)



\bibitem{DiaSha} P. Diaconis and  M. Shahshahani {\it J. Appl. Probab. A}{\bf 31} , 49 (1994)



\bibitem{CMW} C.C. Chamon, C. Mudry, and X.-G. Wen  {\it Phys. Rev. Lett. } {\bf 77},  4194 (1996)

\bibitem{Wid} H. Widom {\it Amer. J. Math.} {\bf 95} 333 (1973)

\bibitem{FWSel} P.-J. Forrester and S.O. Warnaar  {\it Bull. Amer. Math. Soc. (N.S.)} {\bf 45}, 489 (2008)

\bibitem{LLR} M.R. Leadbetter, G. Lindgren, $\&$ H. Rootzen. {\it Extremes and related properties of random sequences and processes.} (Springer-Verlag. New York, 1982)

\bibitem{BZ}   M. Bramson and O. Zeitouni
{\it Comm. Pure Appl. Math.} {\bf 65} 1 (2012); E. Bolthausen et al. {\it Elec. Comm. Prob.} {\bf 16} 114 (2011)

\bibitem{ABK} L.-P. Arguin, A. Bovier, N. Kistler
{\it Comm. Pure Appl. Math.} {\bf 64} 1647 ( 2011) and {\it ArXiv:} 1201.1701

\bibitem{DS} B. Derrida and H. Spohn {\it J. Stat. Phys.} {\bf 51}, 817 (1988); C. Webb {\it J. Stat. Phys.}
{\bf  145}, 1595 (2011)

\bibitem{turbfin} R. Baile and J.-F. Muzy {\it Phys. Rev. Lett.} {\bf 105}, 254501 (2010);  D. Ostrovsky {\it Rev. Math. Phys.} {\bf 23} 127 (2011);
D. B. Saakian et al. {\it EPL} {\bf 95} 28007 (2011)

\bibitem{Hia} G.A. Hiary   {\it Math. Comp.} {\bf 80}, 1785 (2011)



\end{thebibliography}
\end{document}